\documentclass{article}
\usepackage[utf8]{inputenc}
\PassOptionsToPackage{utf8}{inputenc}
\RequirePackage{fix-cm}

\usepackage{algpseudocode,algorithm,algorithmicx}
\newcommand*\Let[2]{\State #1 $\gets$ #2}

\usepackage{url}
\usepackage{graphicx}
\usepackage[dvipsnames]{xcolor}
\usepackage{amsfonts}

\title{Signal metrics analysis of oscillatory patterns in bacterial multi-omic networks}
\author{Francesco Bardozzo $*^1$, Pietro Li\'o $*^2$, Roberto Tagliaferri$*^1$  \\ \\  $*^1$ - DISA-MIS - University of Salerno (IT), \\ $*^2$ - Computer Laboratory, University of Cambridge (UK) }
\date{August 2020}

\usepackage{natbib}
\usepackage{graphicx}

\begin{document}
\maketitle
\abstract{\textbf{Motivation:} One of the branches of Systems Biology is focused on a deep understanding of underlying regulatory networks through the analysis of the biomolecules oscillations and their interplay. Synthetic Biology exploits gene or/and protein regulatory networks towards the design of oscillatory networks for producing useful compounds. Therefore, at different levels of application and for different purposes, the study of biomolecular oscillations can lead to different clues about the mechanisms underlying living cells. It is known that network-level interactions involve more than one type of biomolecule as well as biological processes operating at multiple omic levels. Combining network/pathway-level information with genetic information it is possible to describe well-understood or unknown bacterial mechanisms and organism-specific dynamics.
		\\
		\textbf{Results:} Network multi-omic integration has led to the discovery of interesting oscillatory signals. Following the methodologies used in signal processing and communication engineering, a new methodology is introduced to identify and quantify the extent of the multi-omic oscillations of the signal. New signal metrics are designed to allow further biotechnological explanations and provide important clues about the oscillatory nature of the pathways and their regulatory circuits. Our algorithms designed for the analysis of multi-omic signals are tested and validated on 11 different bacteria for thousands of multi-omic signals perturbed at the network level by different experimental conditions. Information on the order of genes, codon usage, gene expression, and protein molecular weight is integrated at three different functional levels. Oscillations show interesting evidence that network-level multi-omic signals present a synchronized response to perturbations and evolutionary relations along with taxa.
		\\
		\textbf{Availability:} The algorithms, the code (written in R), the tool, the pipeline and the whole dataset of multi-omic signal metrics are available at a GitHub repository:  \url{https://github.com/lodeguns/Multi-omicSignals}\\
		\textbf{Contact:}  robtag@unisa.it

	\section{Introduction}
	\label{intro}
	An oscillating multi-omic network is a complex interlacing of interacting elements, which could be defined gene/protein oscillators. These elements are able to produce oscillations of a certain frequency potentially involving several and different cell processes \citep{govindarajan2012compartmentalization}.
	It is proved that biomolecules, such as genes or proteins, could exhibit oscillatory behaviors. Moreover, if combined, they could generate oscillatory circuits observable on the temporal axis. The gene/protein oscillations are shown to be controlled by underlying regulatory networks and impact to different scales. \citep{michalodimitrakis2008engineering, shis2018dynamics}.  In \cite{levine2013functional} a comprehensive review is reported, showing that, in general, living cells have a pervasive dynamic behaviour where the key transcription and regulatory factors oscillate on and off repeatedly even when cells are in steady states. These oscillations can be detected in circuits of genes/proteins involved in stress responses, signaling, and cell development \citep{lenz2011temporal}. Many biomolecules exhibit network-level metabolic interactions coordinated with cell growth, chromosome replication and cell division \citep{wang2009metabolism}. For example, the Min oscillation is discovered to be fundamental in the E.\emph{coli} cell division \citep{lutkenhaus2008min}. In addition, biomolecule oscillations are found in very complex and multi-periodic signals. \cite{amariei2014quantifying} generate a de-noised waveform from multiple significant frequencies, to provide oscillation statistics including signal metrics and multi-periodicity quantification. Furthemore, genetic circuits present some interesting intrinsic dynamics \citep{guantes2010trade}; for example, in signalling pathways, they are responsive to feedback loops and show functional plasticity \citep{lenz2011temporal}.  Recently, pseudo-temporal estimations based on the level of mRNA and/or proteins have been introduced to detect oscillatory gene networks from single-snapshot experiments \citep{boukouvalas2019osconet}. Unfortunately, even if the complex dynamics of bacterial processes could be predicted by keeping track of bacterial functional adaptation to single-snapshot perturbations  (control vs treatment), the oscillatory dynamics have not yet been explored enough. Thus, a complete mapping of regulatory and control mechanisms is not yet known. Furthermore, the lack of experiments along temporal axis make difficult to recognise oscillating biomolecules and their circuital interactions. \citep{prokop2013systems}. However, network-level synchronisation could be outlined by the hypotheses that every biomolecule in a network could interact with any other; this causes that the shared biomolecule oscillators synchronise the signals on common fluctuations. There are few examples in nature for which these assumptions are fully verified, such as \emph{circadian oscillations} \citep{golden2003timekeeping}. On the other hand, in synthetic biology, \emph{artificial oscillators}, although often showing poor accuracy \citep{potvin2016synchronous, hawe2019inferring}, are one of the most promising research fields. In particular, \emph{artificial oscillators} allow the creation of genetic circuits focused on the execution of logical programming in living cells. The E.\emph{coli} \emph{repressillator} experiment represents a clear example of how genetic regulatory networks can be designed and implemented to perform new functions \citep{arenas2008synchronization, elowitz2000synthetic}. In our previous works \citep{bardozzo2015multi, bardozzo2018study}, we investigated the E. \emph{coli} response to $\approx$ 70 perturbations by monitoring network-level oscillation changes from controls to treatments and discovering that at the network level, there is another type of \emph{inter molecular} multi-omic oscillation associated to each single pathway and experiment.  This type of oscillation, as far as we know, has not yet been sufficiently investigated on fixed-time analysis. In particular, our multi-omic oscillations could be described as a multi-periodic signal given by the \emph{variation} of interacting biomolecular multi-omics. This \emph{variation} should be intended in terms of sequence low-hight alternations of multi-omic values \citep{bardozzo2018study}. To clarify these points, in Section \ref{ss1} and Figure \ref{Fig1}, the identification procedure of these multi-omic signal is described.  With respect to our previous works, the number of organisms is extended to 11. Also, the experiment cardinality has increased to the order of thousands. Previously, we demonstrated how to measure the structural relations between the genomic and proteomic layers and how these led to oscillatory variations in response to perturbations. On the contrary, in this paper, taking advantage of signal theory and communication engineering, \emph{ad hoc} metrics to better quantify network-level oscillatory features are designed and the algorithms for their computation are provided on an online repository. In detail, two new change point detection algorithms ($CPD$) \citep{siegmund2013change, unakafov2018change} are introduced. These algorithms are capable of managing both the complexity of the variable amplitude of the multi-omic signal and the multiple periodicity.  Further analyses on network-level synchronisations based on our novel signal metrics are provided. In particular, the analyses are focused on the interactions through different pathways of the same organism and modulated by different condition contrasts (\emph{CC}) (single-snapshot $mRNA$ experiments) \citep{meysman2013colombos}. Through our approach, it is possible to recognise the oscillating networks, eventually evaluating if they are synchronised (periodically and simultaneously activated) or not synchronised, and how this feature changes across taxa. Integrated multi-omics are created from the following single omics:  the codon usage \citep{sharp1987codon}, mRNA amount contrasts \citep{meysman2013colombos} and the protein molecular weight.  Our final dataset is composed of $2.830.722$  multi-omic signals from 11 different bacteria on thousands of environmental experiments. Network-level oscillatory variations are analysed with three functional levels of granularity from KEGG orthology \citep{kanehisa2000kegg}.   The results confirmed and extended our previous findings, by showing that network-level multi-omic oscillations exist in bacteria. Moreover, we found additional clues to support that the oscillatory networks are synchronised showing a combined dynamic response to perturbations. Furthermore, the comparisons between the various bacteria succeed in highlighting, in a completely innovative way, how network-level oscillations could reflect the effects of evolutionary pressure. Moreover, even if the maintenance of the gene order is not well understood \citep{rocha2003dna}, this research could give new clues to its meaning, underlining its dynamical relations with the proteomic layer under the evolutionary pressure \citep{tmes2001evolution}.

	\begin{figure}[!t] 
		\includegraphics[width=\textwidth]{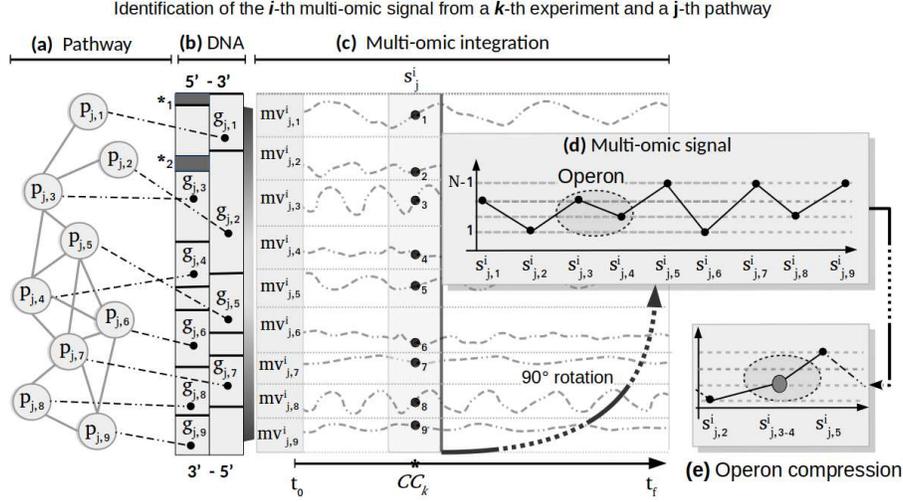}
		\caption{The multi-omic information concerning one of the 11 bacteria is collected and data-integrated with respect to the \emph{j}-th pathway and considering the \emph{k}-th perturbation (indicated as the \emph{k}-th experiment, or better condition contrast $\mathbf{CC_k}$). As it is described in Section \ref{defnorm} and shown in Box \textbf{(a)} and Box \textbf{(b)}, the multi-omic information is combined and grouped by molecular networks (KEGG pathways) and ordered with respect to the gene order information. The dashed lines between pathway nodes and genes indicate a \textbf{1:1} correspondence between the $n$ selected pathway proteins and their related genes ($  \forall i \in [1,n], p_{j,i} \Leftrightarrow g_{j,i} $).	As shown in \textbf{Box (c)}, the multi-omic values ($\vec{mv}^i_j$) are represented as outputs of gene/protein oscillators at time $t_0 \leq k \leq t_f$,  which is the time when a single condition contrast is taken. Then, the ($\vec{mv}^i_j$)  are discretized into \textbf{N} quantisation levels (see also Section \ref{ioac}), and rotated of $90^{\circ}$ with respect to the multi-omic space. In this way, as it is shown in Box \textbf{(d)}, a discrete multi-omic signal $\mathbf{s^i_j}$ is obtained, which could be eventually operon compressed (see Box \textbf{(e)} and also Section \ref{oc}). The $\mathbf{*^1}$ and $\mathbf{*^2}$ asterisks represent the proximal and the distal positions of possible promoters or repressors neighbouring the operon $\mathbf{g_{j,3}},\mathbf{g_{j,4}}$. }\label{Fig1}
	\end{figure}

		\section{Methods}
		\label{S:2}
		In subsection \ref{ss1} the procedure for multi-omic signal identification is described. In Figure \ref{Fig1} an overview of this task is provided.  In detail, the mRNA condition contrasts and protein weight information are extracted, normalised and aligned with the codon usage information. In Figure \ref{Fig1} - Box \textbf{(c)}, the multi-omic signal identification in condition contrast $\mathbf{CC}$  at a fixed time $\mathbf{k}$ is indicated. From this single experiment, the collected multi-omic values $\vec{mv}$ are selected with respect to the KEGG pathway composition (Box \textbf{(a)}) and ordered considering the related DNA spatial positions (Box \textbf{(b)}). Then, they are figuratively \emph{rotated} of $90^{\circ}$ and normalised generating a network-level multi-omic signal (Box \textbf{(d-e)}). As shown in \textbf{Box (d)}, the multi-omics are discretised by considering \textbf{N} common levels of discretisation obtained through a between-organisms analysis (\emph{IOAC} procedure) as defined in subsection \ref{defnorm}.The identified signals have been found to be almost periodic, then in subsection \ref{ss2}, two novel algorithms are introduced for their analyses. Furthermore, by applying Algorithm \ref{alg:algo1} across the organisms and for all the multi-omic combinations, the periodicity in the multi-omic signals is identified through an estimate $\hat{\theta}$ of the search window where it can be found (Figure \ref{Fig2} - Box \textbf{(a)} and Section\ref{subs1}). Next, another CPD algorithm (Algorithm \ref{alg:algo2}) is applied to obtain two signal indices: $osc_s$ and $osc_k$ (Figure \ref{Fig2} - Box \textbf{(b)}).   In particular, they are used to identify network-level oscillations for each pathway, for each experiment and for each multi-omic combination (Figure \ref{Fig3}- Box \textbf{(a)}). Furthermore, the oscillations between networks and among the experiments, as described in the pipeline of Figure \ref{Fig3}- Box \textbf{(a-b-c)} are computed within and between organisms. Finally, this step is well described in Section \ref{netsync}.

		\subsection{Multi-omic signals}
		\label{ss1}

		\subsubsection{Definition and normalisation }
		\label{defnorm}
		Given $ \vec{P_{j,n}}$ as the \emph{j}-th bacterial pathway of \emph{n} proteins, the multi-omic signal $s_j$ is composed by a finite vector of \emph{n} multi-omic values \textbf{$\vec{mv_j}$} associated to a subset of genes $ \vec{G_{j,n}} =\lbrace g_{j,1}, g_{j,2}, \dots g_{j,n} \rbrace $. The $g_{j,i}$ are collected in  $ \vec{G_{j,n}} $  considering the exact correspondence \textbf{1:1} with respect to the proteins $p_{j,i}$  that compose the $ \vec{P_{j,n}} = \lbrace p_{j,1}, p_{j,2}, \dots p_{j,n} \rbrace $.	Each multi-omic value \textbf{$\vec{mv_{j,i}}$} is arranged on the multi-omic signal $s_{j,i}$ considering the relative position of its associated gene $g_{j,i}$ with respect to the origin of replication.   In order to describe the signal we adopt the form $s_j[k]$ with the index $k \in \mathbb{Z}$ (such for example, the \emph{i}-th multi-omic value of a signal is equal to $s_j[i] = mv_{j,i}$).
		The multi-omic values $\vec{mv_{j,i}}$ are combined averaging their associated single-omics $\vec{sv_{j,i}}, \forall i \in \mathbb{Z}$ .
		Since the  $\vec{sv_{j,i}}$ are not defined in the same range, a normalisation is applied to make their values comparable.  In particular the single-omics are three: \textbf{(I)} the mRNA condition contrasts (\emph{CC}) and \textbf{(II)} the molecular weights (MW) and  \textbf{(III)} the codon adaptation index. MW and \emph{CC} are normalised into the interval [0,1], that is the same range in which the codon adaptation index (CAI) is already defined \citep{sharp1987codon}. In section \ref{subs2} there is an accurate description of these sources.
		
		\subsubsection{Multi-omic discretisation through the IOAC procedure }
		\label{ioac}
		
		In order to compare the signals between different organisms, an amplitude discretisation process was applied. The whole procedure is called: Inter-Organisms Amplitude Consensus (IOAC) and discretisation. The signal amplitude is divided into \textbf{N} bins ($C=\lbrace 1,2,..b-1, b, b+1...N \rbrace, \forall b \in \mathbb{Z} $) and the original $\vec{mv_{j,i}}$ was replaced by the bin label it belongs to through a function map:  $f_m(s) := s[i] \rightarrow  c(s[i]), \forall i \in \big[1,n\big] $. Thus, if the total number of classes (\textbf{N}) is equal to the cardinality of $ C $, then each class $ c(s[i]) $ represents the $c$-th interval in which the $\vec{mv_{j,i}}$ falls. However, the correct estimation of \textbf{N} for the discretisation follows a study on the single omic distributions trough the IOAC. In particular, the distribution of the $\vec{sv_{j,i}}$ on the whole genome and for each organism was investigated by means of the Anderson-Darling test (A-D Test, \citep{razali2011power}). In the case of CAI, for the $91.67\%$, the A-D test rejects $H_0$ with a significance of $0.05$ . The MW and \emph{CC} have the same significance with a percentage near to $100\%$. We conclude that the single omics do not follow a normal distribution. As a consequence, the optimal number of \textbf{N} bins is computed for non-normal distributions applying the Doane’s formula \citep{venables2013modern}. The N values are estimated between-organisms. In particular, \textbf{N} is fixed equal to 9 troughs a frequency based consensus (bold column in Supplementary Material Section 1).
		Consequently, in our set-up, the multi-omic signals generated for each experiment are discrete non-deterministic signals that represent gene-ordered multi-omic values that fall into 9 possible class intervals (from 0 to 8).
		\subsubsection{Operon compression }
		\label{oc}
		The extent of the multi-omic signal dataset is increased by their operon compressed versions. In this case, if a set of multi-omic values in a signal $s_j$ of length $n$ are part of an operon in position $r$ of length \emph{m}, this set is defined as $s_j[r:r+m] = \lbrace mv_{j,r}, mv_{j, r+1}, mv_{j, r+m} \rbrace, with \ m<n$. In this case, each $mv_{j,i}$ follows its natural adjacent disposition on the DNA sequence. For this reason, we can represent the signal as a concatenation (indicated as $\oplus$) of the original signal with respect the operon: $s_j[1:n] = s_j[1:r-1] \oplus s_j[r:r+m] \oplus s_j [n-m+r:n]$. According to their natural regulatory functions, in order to apply the compression, the $mv_{j,i}$ that composes an operon could be seen as a single averaged value : $ s_j[r] = f_m(|mv_i + mv_{i+1} + \dots + mv_{m} / m|)$. In this way we obtain additional signals with this shape: $s_j[1:n] = s_j[1:r-1] \oplus s_j[r] \oplus s_j[n-1+r:n]$. Obviously, the compression is applied more times if  occurs, thus shortening the signal length.

		\subsection{Multi-omic oscillation analysis}
		\label{ss2}

		\subsubsection{Variable half-periods estimation with a change-point detector }
		\label{subs1}
		In this section, we deal with multiple-periodicity signals characterised from different \emph{mRNA} condition contrasts (see Section \ref{subs2}). These could represent more replications of the same experiment. Thus, due to the experimental intrinsic and extrinsic noise \citep{singh2013quantifying}, it is very rare that these signals follow an ideal shape with a fixed periodicity; on the contrary, the periodicities are more variable making difficult the oscillation detection. After all, if there are oscillations, then their half-periods (from peak to lows or vice versa) are localised in windows of variable length. In order to obtain an estimation of half-periods, we introduce a novel localised and non-parametric change-point detection algorithm (CPD). The idea behind Algorithm  \ref{alg:algo1} is based on the analysis of the median multi-omic variations estimating the window lengths $\theta$s in which the half-periods occur. A change-point is detected if and only if the median value of the previous variations ($\vec{m_k}$) is less or equal to the new coming multi-omic variation ($d_k$) respecting the genes order. In general, the change point detectors look for changes in the statistical characteristics of the signal (i.e. the median values $\vec{m_k}$), therefore considering the signal as a collection of different distributions arranged in adjacent windows \citep{darkhovski1994nonparametric}.
		In Figure \ref{Fig2} - \textbf{Box (a)} an atomic example of a half-period length estimation ($\theta$) is shown. Algorithm \ref{alg:algo1} collects in a vector $\vec{ \theta }$ all the $\theta$s computed along the signal. Then, for each signal, the median values of $\vec{ \theta }$ are collected. Next, a common $\hat{\theta}$ is defined as the maximum between the median values of $\vec{ \theta }$ between all the organisms and for each multi-omic combination.
		In particular, the $\hat{\theta}$s have a double functionality; they are indices of the different multi-omic interplay within and between organisms and they are halt condition parameters of Algorithm \ref{alg:algo2}.  In our case, we are focusing on the variable half-period lengths for all the pathways. It is discovered that all the organisms, on overall experiments and for each pathway, show a median $\theta$ of about 3.0 and an average comprised between 3.0 and 3.7 with a low standard deviation. Also, the max and min  $\theta$s values are very similar between organisms.  These statistics depend on the different multi-omic combinations (\textbf{MOC}) and on the presence of operon compression. The table of the $\hat{\theta}$s is shown in the Supplementary Material Section 8, while in Section 2 the source code of Algorithm \ref{alg:algo1} is provided.

		\begin{figure}[!b]
			\includegraphics[scale=0.25]{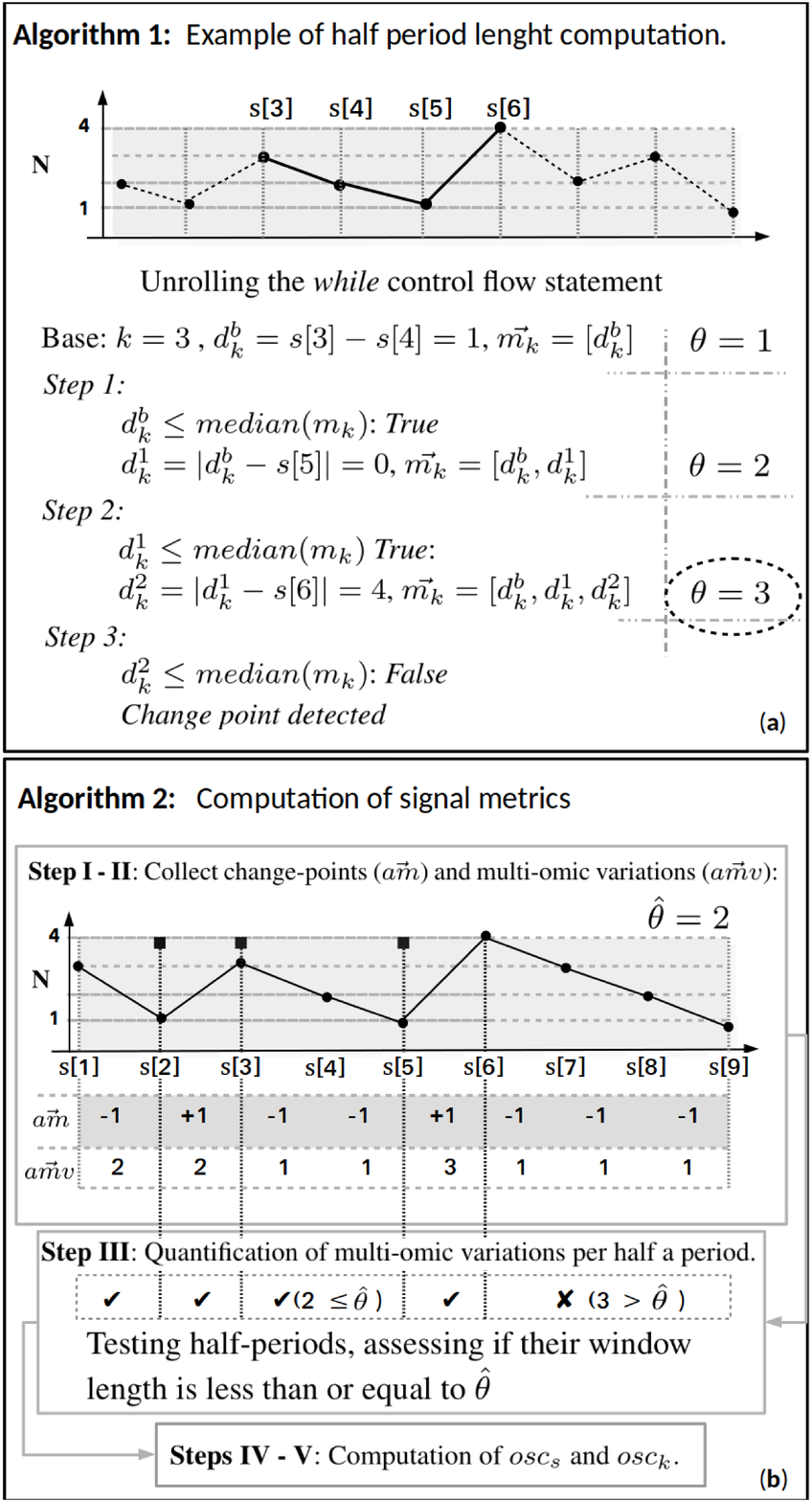}
			\caption{ Algorithm \ref{alg:algo1} and Algorithm \ref{alg:algo2} are applied to identify the periodicity of 2,830,722 signals. An atomic example of half-period estimation is shown in \textbf{Box (a)}. In this case, a half-period is recognised with a length of $\theta =3$. In general, Algorithm \ref{alg:algo1} collects in a vector $\vec{ \theta }$ all the $\theta$s computed along the signal. The  $\hat{\theta}$ are computed between organisms as the max of the half-period median lengths of $\vec{\theta}$.   The $\hat{\theta}$s have the same value in all the organisms but vary depending on the multi-omic combination considered with or without operon compression (see Section \ref{subs1}. Next, Algorithm \ref{alg:algo2} is divided into a 5-step pipeline as described in Section \ref{algo2}. In \textbf{Box (b)}, these steps are summarised. Algorithm \ref{alg:algo2} takes in input a signal and its associated $\hat{\theta}$. Then, in \textbf{Step I-II} both adjacent change-points and multi-omic variations are computed along the signals. In \textbf{Step III} the adjacent multi-omic variations ($amv[p^e], amv[p^{e+1}], \dots ]$) are added together for each window until a halt condition occurs. In \textbf{Box (b)-Step III}, the halt condition is represented by $p^{e+i} - p^{e}=3$ greater than $\hat{\theta}$. Finally, in \textbf{Step IV-V}, Algorithm \ref{alg:algo2}  gives in output the two metrics $osc_s$  and $osc_k$. All the variables shown in the Figure are defined and described in the respective sections.   }\label{Fig2}.
		\end{figure}
		\begin{algorithm}
			\caption{Multi-omic median window periodicity with CPD \label{alg:algo1}}
			\begin{algorithmic}
				\Require{ A multi-omic signal: $s[n]$ of length $n$ and \textbf{N}  the between organisms bins estimation}
				\Function{change-point-median-window}{$s[n], N$}
				\Let{$\vec{\theta}$}{$NULL$} \Comment{ Array of enstimated half-period lenghts $\theta$s }
				\For{$k \gets 1 \textrm{ to } n$}
				\Let{$\theta$}{1} \Comment{Index of the current window.}
				\Let{$d_k$}{$|s[k] - s[k+1]|$}
				\Let{$\vec{m_k}$}{$[d_k]$} \Comment{Trace the multi-omic variation}
				\If{$k+1 < n$}
				\While{$d_k \leq median(\vec{m_k})$}  \Comment{Change-point detection}
				\If{k+2 < n-1}
				\Let{$d_k$}{$|d_k - s[k+2]|$ }
				\Let{$k$}{$k+1$}
				\Let{$\vec{m_k}$}{$[\vec{m_k} \oplus d_k]$}
				\EndIf
				\Let{$\theta$}{$\theta +1$}
				\EndWhile
				\EndIf
				\Let{$\vec{m_k}$}{$NULL$}
				\Let{$\vec{\theta}$}{$[\vec{\theta} \oplus \theta]$}
				\EndFor
				\State \Return{$\vec{\theta}$}
				\EndFunction
			\end{algorithmic}
		\end{algorithm}

		\subsubsection{Multi-omic signal indices: $osc_s$ and $osc_k$}
		\label{algo2}
		In this section, we introduce algorithm \ref{alg:algo2}, which is a multi-level change point detector, capable of detecting multi-omic variations in relation to the half-periods to which they belong. The algorithm returns in output  $osc_s$ and $osc_k$. In particular, $osc_s$ is an oscillation index which relates the length of the half-periods (conditioned by the multi-omic variations) with their signal amplitude. Instead, $osc_k$ is an index describing the relative length of the half-periods with respect to the signal length. The algorithm is divided into five steps, as they are summarised in Figure \ref{Fig2} - Box \textbf{(b)} and detailed as follows: \\
		\textbf{I) Collect multi-omic adjacent change-points: } In this step algorithm \ref{alg:algo2} collects the adjacent multi-omics change-points on the signal $s$ tracking the adjacent multi-omic class variations $\vec{am}$ as described in Equation \ref{eq:01}:
		\begin{equation}
		\vec{am[i]} \leftarrow \Bigg\{ \begin{tabular}{ccc}   +1 & if & s[i] > s[i+1] \\   0  & if & s[i] = s[i+1] \\   -1 & & otherwise  \end{tabular} \ \ \ \ \forall i \in |s| \ . \label{eq:01}
		\end{equation}
		\textbf{II) Collect adjacent multi-omic variations: }\\
		For each signal $s$, algorithm \ref{alg:algo2} collects the adjacent multi-omic variations $\vec{amv}$ as the absolute difference between two adjacent multi-omic values, as described in Equation \ref{eq:02}:
		\begin{equation}
		\vec{amv[i]} \leftarrow \ |s[i] - s[i+1]| \ \ \ \ \forall  i \in |s| \ . \label{eq:02}
		\end{equation}
		\textbf{III) Quantification of multi-omic variations per half-periods: }\\
		The multi-omic variations of $\vec{amv[i]}$ are summed from a starting point ($p_s^e$) to a halt point ($p_h^e$) at time step  $e$. The procedure is repeated iteratively  from a $p_h^e$ to another halt point until the length of the signal is reached. The results are given in output in the vector $\vec{mvq}$, as shown in Equation \ref{eq:03}:
		\begin{equation}
		\vec{mvq} = \Big[  \sum_{i= p_s^e}^{p_h^e}amv[i], \sum_{i=p_h^{e}}^{p_h^{e+1}}amv[i], \ \  \ \dots  \ \ ,    \sum_{i=p_h^{e+n-1}}^{p_h^{e+n}}amv[i]\Big] \ .\label{eq:03}
		\end{equation}
		Therefore, the last $p_h^{e-1}$ become the new $p_s^{e}$.  In particular, the halt point is computed dynamically with two stop conditions: the former is given when a change point $\vec{am[i]} \neq 0$, the latter is given when $p_h^e-p_s^e \geq \hat{ \theta }$.  The number of the summed multi-omic variations for each step $e$ are collected in the vector  $\vec{mvl}$ as shown in Equation \ref{eq:04}.
		\begin{equation}
		\vec{mvl} = \Big[ |p_h^{e+t} - p_s^{e +t}| \Big], \forall t \in |s| \ . \label{eq:04}
		\end{equation}
		The calculation of $\vec{mvq}$ and $\vec{mvl}$  is the central point for the development of our change point detector, therefore \textbf{Step III} has been described in detail in the pseudocode of Algorithm \ref{alg:algo2} and is illustrated in the example of Figure \ref{Fig2} for an halt condition.
		
		\textbf{IV) Computation of $osc_s$: }\\
		The two vectors $\vec{mvq}$  and the $\vec{mvl}$ have the same length $d$, where at most $d \leq |s|-1$. The first one traces the quantified multi-omic variation for each half-period. The second one traces the window lengths in which the variations are computed. Thus, the oscillation index is defined in Equation \ref{eq:05} as the product of the two vectors with respect to the sum of $\vec{mvl}$ for the \textbf{N} bins:
		\begin{equation}
		osc_s =  \frac{\sum_{i=1}^{d} \vec{mvq[i]} \ \cdot \ \vec{mvl[i]}}{ (\mathbf{|N|}-1)\sum_{1}^{d}\vec{mvl} } \ . \label{eq:05}
		\end{equation}
		The $osc_s$ is defined in the interval $[0,1]$. If the signal oscillation is null it returns 0, if it is perfect returns 1. The intermediate values of the $osc_s$ are indices of the signal oscillations. In order to give a proof of the algorithm correctness, we prove the following theorem and some associated corollaries, as deepened in the Supplementary Materials Section 3.
		\emph{\textbf{Theorem I}: Algorithm \ref{alg:algo2} gives in output an oscillation index $osc_s$ equal to 1 if and only if ($\Leftrightarrow $) the observed signal presents a perfect oscillation.} \\ \\
		\textbf{V) Computation of $osc_k$:}\\
		Algorithm \ref{alg:algo2} computes, also, the oscillation index $osc_k$ in Equation \ref{eq:06}: \begin{equation}
		osc_k = \frac{|\vec{mvl}|}{|s|} \ . \label{eq:06}
		\end{equation}
		For each signal, this index represents a relation between the length of $s$ and the number of periods along the signal, described as the cardinality of $\vec{mvl}$. As we will see, $osc_k \in [0,1]$ remains defined in a certain interval and describes some interesting relations in the analysis of the pathways phase synchronisations (Section \ref{results}).   In the Section 5 of Supplementary Material we provide the source code related to Algorithm \ref{alg:algo2} in order to compute $osc_s$ and $osc_k$.

		\begin{algorithm}
			\caption{\label{alg:algo2}: Oscillation indices: $osc_s$ and $osc_k$. The procedure is divided into 5 steps as described in Section \ref{algo2}. Here, in pseudocode, Step III in relation with the other steps is shown.}
			\begin{algorithmic}
				\Require{ A multi-omic signal: $s[n]$ of length $n$ and estimated $\hat{\theta}$ }
				\Function{compute-$\vec{mvq}$-and-$\vec{mvl}$ }{$s[n], \hat{\theta}$}
				\Let{$\vec{am}$}{Step I} \Comment{ Collect multi-omic adjacent change-points.}
				\Let{$\vec{amv}$}{Step II} \Comment{ Collect adjacent multi-omic variations.}
				\Let{$p_s$}{0}
				\Let{$p_h$}{0}
				\Let{$j$}{0}
				\For{$e \gets 1 \textrm{ to } n$}
				\Comment{Change-point detection clauses: $c_1$, $c_2$}
				\Let{$c_1$}{$am[e] < 0 \wedge am[e-1] > 0 $}
				\Let{$c_2$}{$am[e] > 0 \wedge am[e-1] < 0$}
				
				\If{$c_1 \vee c_2$}
				\Let{$p_h^e$}{$e$}
				\Let{$\vec{mvq}[j]$}{$\sum_{i= p_s^e}^{p_h^e}amv[i]$}
				\Let{$\vec{mvl}[j]$}{ $|p_h^{e} - p_s^{e}|$}
				\Let{$j$}{$j+1$}
				\Let{$p_s^e$}{$e$}
				\Else
				\Comment{No-change-point detection clauses: $c_3$, $c_4$}
				\Let{$c_3$}{$am[e] \leq 0 \wedge am[e-1] \leq 0 $}
				\Let{$c_4$}{$am[e]  \geq 0 \wedge am[e-1] \geq 0 $}
				\If{$c_3 \vee c_4$}
				\If{$(e -  p_s^e) < \hat{\theta}$}
				\Let{$e$}{$e+1$}
				\Else
				\Let{$p_h^e$}{$e$}
				\Let{$\vec{mvq}[j]$}{$\sum_{i= p_s^e}^{p_h^e}amv[i]$}
				\Let{$\vec{mvl}[j]$}{ $|p_h^{e+t} - p_s^{e +t}|$}
				\Let{$j$}{$j+1$}
				\Let{$p_s^e$}{$e$}
				\EndIf
				\EndIf
				\EndIf
				\EndFor
				\State \Return{$(\vec{mvq},\vec{mvl})$}
				\EndFunction
				\Let{$osc_s$}{Step IV} \Comment{Compute the oscillation index $osc_s$.}
				\Let{$osc_k$}{Step V} \Comment{Compute the oscillation index $osc_k$.}
			\end{algorithmic}
		\end{algorithm}

		\subsubsection{Robustness and sensitivity analysis of Algorithm \ref{alg:algo2}}
		\label{robsen}
		The robustness of Algorithm \ref{alg:algo2} was tested by defining two types of perturbations.
		Without loss of generality, we assumed that the perturbations are defined by a random distribution with zero mean and unit variance. As a consequence, the first type of perturbation applied to the discrete signal \emph{s} is a stochastic additive noise. We decided to add the $5\%$ of the generated noise, in the following way: $s[i] + (\mathcal{N}(0,1) * 0.5) \ \forall i \in [1, n]$. The second type of perturbation consists of random shuffling the elements of the original signal, thus testing the importance of the information deriving from the gene order. The t-test p-value of the obtained $osc_s$ on random shuffled distributions is equal to $0.004385$, while on random additive noise distributions it is less than $2.2e-16$. We selected a subset of original signals with at least  $70\%$ of significant oscillating multi-omics. This means that we selected only the signals with oscillation index $osc_s \geq \phi$, with $\phi =0.7$. Then, we perturbed this subset of signals and we newly computed the $osc_s$. In Figure \ref{Fig4}, the PDF of the original signal oscillation index $osc_s$ (solid line) against the perturbed ones (dashed lines) are shown. In particular, we can observe that the $osc_s$ of the original signals remains into the interval from 0.7 to 1, while the noisy and shuffled signals intercept an interval from 0.4 to 1,  thickening the area of interest  (Figure \ref{Fig4}  - $*^2$, $*^3$) to lower values than those defining the original area (Figure \ref{Fig4}  - $*^1$). Note that Algorithm \ref{alg:algo1} and Algorithm \ref{alg:algo2} have linear complexity \emph{\textbf{O(n)}} over the signal length $n$. As expected, small variations in multi-omic values or a random arrangement clearly lower the oscillation index $osc_s$.
		
        \begin{figure}[]
			\centerline{\includegraphics[width=80mm]{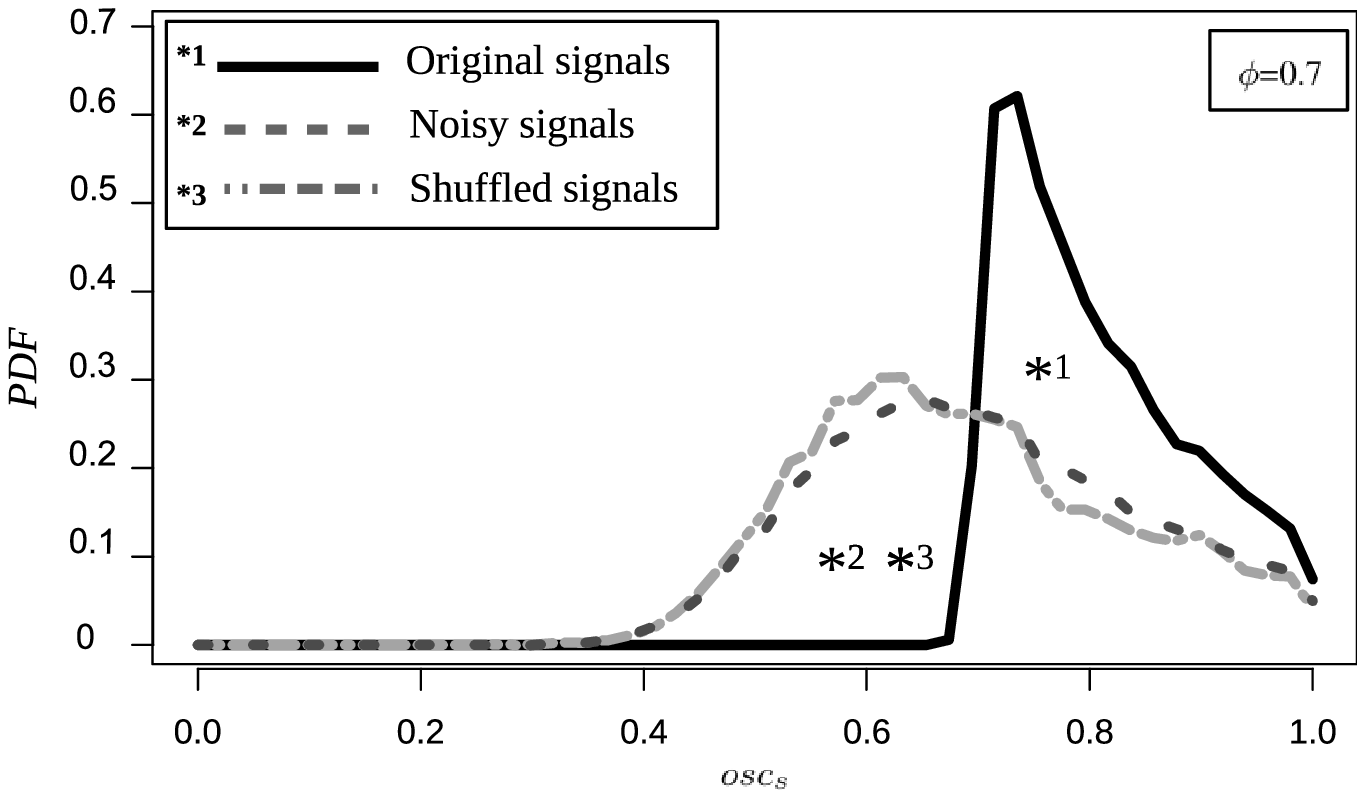}}
			\caption{In this Figure three probability density plots (PDF) are shown. The area under the three PDFs are indicated with the asterisks $*^1, *^2,*^3$. The values of the density functions are on the y-axes. The oscillation index $osc_s$ are on the x-axes. It is plotted the PDF of the original multi-omic signals with $osc_s \geq \phi $ with $\phi$=0.7 (solid line). As it is proved, the underlying area $*^1$ is comprised into the interval of $osc_s$ from 0.7 to the upper bound of 1.0. Then, this subset of original signals ($osc_s \geq 0.7 $) is perturbed in two ways. The PDF of the original signals perturbed with noise are shown as dashed lines and the underlying area is indicated with $*^2$. Those random shuffled are shown in 3 dots dashed lines and the underlying area is indicated with $*^3$. It is possible to observe that, when the original signals are perturbed, the PDFs area $*^3$ and $*^2$ move mostly on $osc_s$ values comprised between 0.4 and 0.7. Thus, the perturbed signals lower the $osc_s$ proving the Algorithm \ref{alg:algo2} correctness, in terms of robustness and sensitivity analysis. The $osc_s$  is computed for all the signals with a length of at least 6.}\label{Fig4}
		\end{figure}

	     \begin{figure}[!t]
			\centering
			\includegraphics[width=\textwidth]{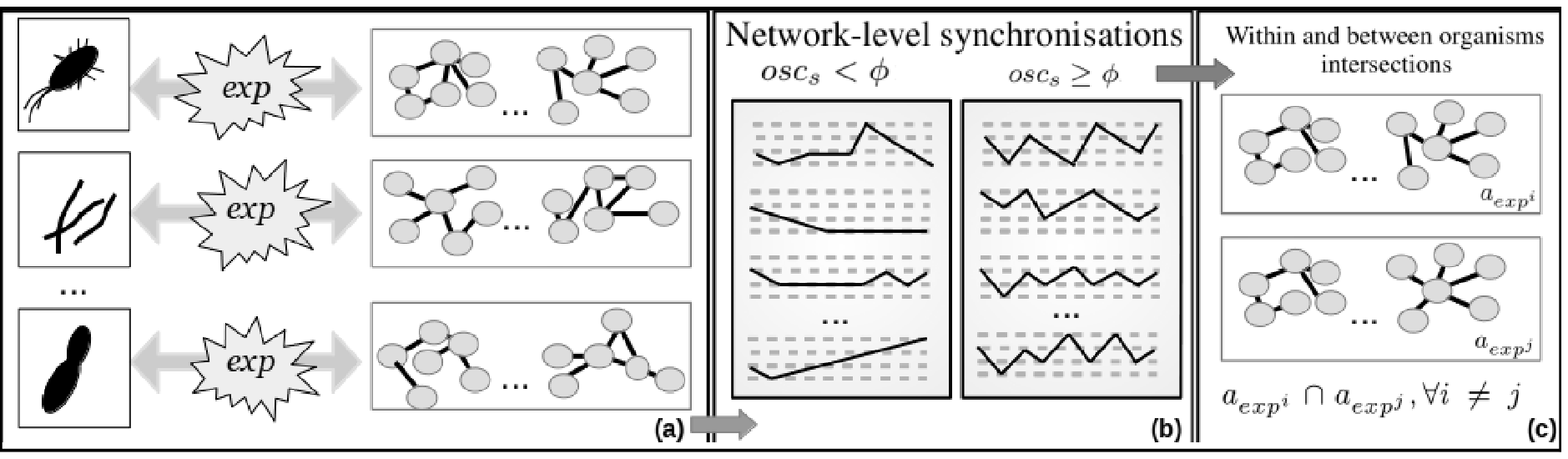}
			\includegraphics[width=\textwidth]{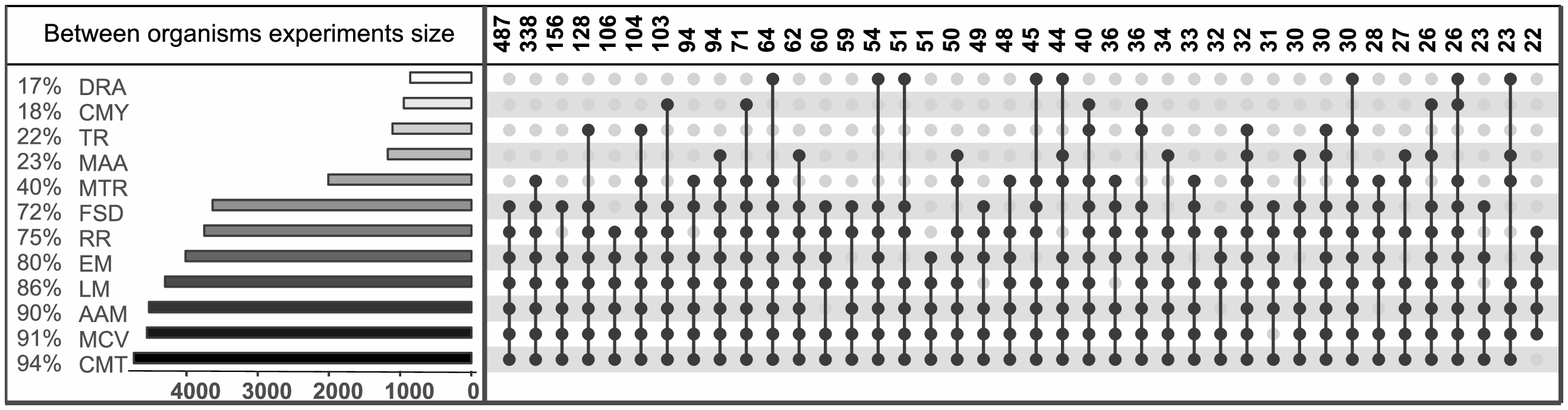}
			\caption{In Figure \ref{Fig3} - Box \textbf{(a-b)} the pipeline of the oscillation extraction is shown. In particular,in Box \textbf{(a)} multi-omic networks are extracted, while in Box \textbf{(b)} the oscillations with a high score ($osc_s \geq \phi, \phi=0.8$) are selected as active pathways and considered as synchronised. In Box \textbf{(c)},
				for each $i$-th experiment the synchronised active pathways are superimposed on those of the $j$-th experiment and their intersections are computed (common activation scheme) with respect to \textbf{KO Level 1}(pathways with the same functions). In particular, for the entire collection of experiments, intersections are calculated within organisms and between organisms.
				In Box \textbf{(d)} an example of between organisms pathway-level synchronisation scheme. Their intersection schemes are reorganised by their \textbf{KO Level 2} functionalities as follows: DRA: \emph{Drug resistance: Antimicrobial}, CMY: \emph{Cell motility},TR : \emph{Translation}, MAA : \emph{ Metabolism of other amino acids}, MTR: \emph{ Membrane transport}, FSD: \emph{Folding, sorting and degradation}, RR: \emph{ Replication and repair}, EM: \emph{ Energy metabolism}, LM: \emph{ Lipid metabolism}, AAM: \emph{ Amino acid metabolism}, MCV: \emph{ Metabolism of cofactors and vitamins}, CMT : \emph{ Carbohydrate metabolism}. The black dots represent the activation scheme that is shared across the between-organism experiments. For example, in the first column we can find 487 between-organism experiments whose the turned-on-simultaneous functionalities are  FSD,RR,EM,LM,AAM,MCV,CMT. This scheme is the most frequent in the between-organism experiments as suggested by the horizontal bars.
				The coverage percentage of the first-column scheme is very significant and overlays with at least the $72\%$ of the detected schemes. }\label{Fig3}
		\end{figure}

		\subsection{Detection of network-level synchronisations }
		\label{netsync}
		The multi-omic oscillation indices:  $osc_s$  and $osc_k$, for all the pathways $\forall P_j \in \mathbf{O}$ on the whole collection of COLOMBOS v3.0 condition contrasts (\emph{CC}s) are computed. We summarised the pipeline of this section in Figure \ref{Fig3}.
		Having set $\phi= 0.8$, the pathways are separated from the others by splitting those with an oscillation index $osc_s < \phi $ from those with $osc_s \geq \phi$.
		A binary function $a$ on $P_j$ is designed defining the pathways with oscillatory behaviours greater than $\phi$ as active pathways ($a(P_x) =1 $), and those less than the threshold as inactive pathways $a(P_y) =0 $, with  $x \neq y$. The active ones are sets of pathways with very relevant oscillatory behaviours. Our hypothesis is that there is a network-level synchronisation only if the $r$ pathways are all active in the same i-th experiment $a_{exp^i} := \lbrace P_{1,i}, P_{2,i}, \dots, P_{r,i}  \rbrace$.  On the other hand, the asynchronised pathways are those that, in the same experiment, are inactive ($P_{j,i} \not\in a_{exp^i}$). For each experiment, the $a_{exp^i}$  are grouped with 3 levels of functional granularities, following their KEGG orthology (\textbf{KO}) classifications, by KEGG pathway names ( \textbf{KO Level 1}), KEGG molecular network functionalities (\textbf{KO Level 2}) and  KEGG maps (\textbf{KO Level 3}). Without loss of generalisation, the rows of \emph{CC}s that represent the same within-studies microarray replications are merged. In this way it is possible to quantify the presence of oscillatory networks on the whole microarray experiment and not only on one of its replications.  The next step in the pipeline consists of an analysis within and between-organisms of the co-occurrence matrices, through the 3 \textbf{KO} levels, in order to understand if the synchronised pathways appear as common scheme overall the experiments ($a_{exp^i} \cap a_{exp^j} , \forall i\not = j $) and to what extent  (see also Figure \ref{Fig3}).  Note that the between-organisms cardinality of the experiments is not homogeneous and it depends on the collection of \emph{CC}s provided by COLOMBOS v3.0. Thus, in Figure \ref{Fig5}, box \textbf{b}, in order to carry on the information about the synchronisation as much as possible, under the heatmap, the cardinality of the experiments (\textbf{EC}) and their relative representation percentage on the between-organisms intersections scheme (\textbf{EE}) are underlined.
		The complete co-occurrence matrices with a fixed threshold to $\phi = 0.8$, their circuital intersections, for each organism and for every multi-omic combinations are provided in the Supplementary Materials - Section 7. A tool capable to visualise these scheme varying the threshold and the other parameters is provided as Supplementary Material Section 7.

		 \begin{figure}[]
			\centerline{\includegraphics[width=90mm]{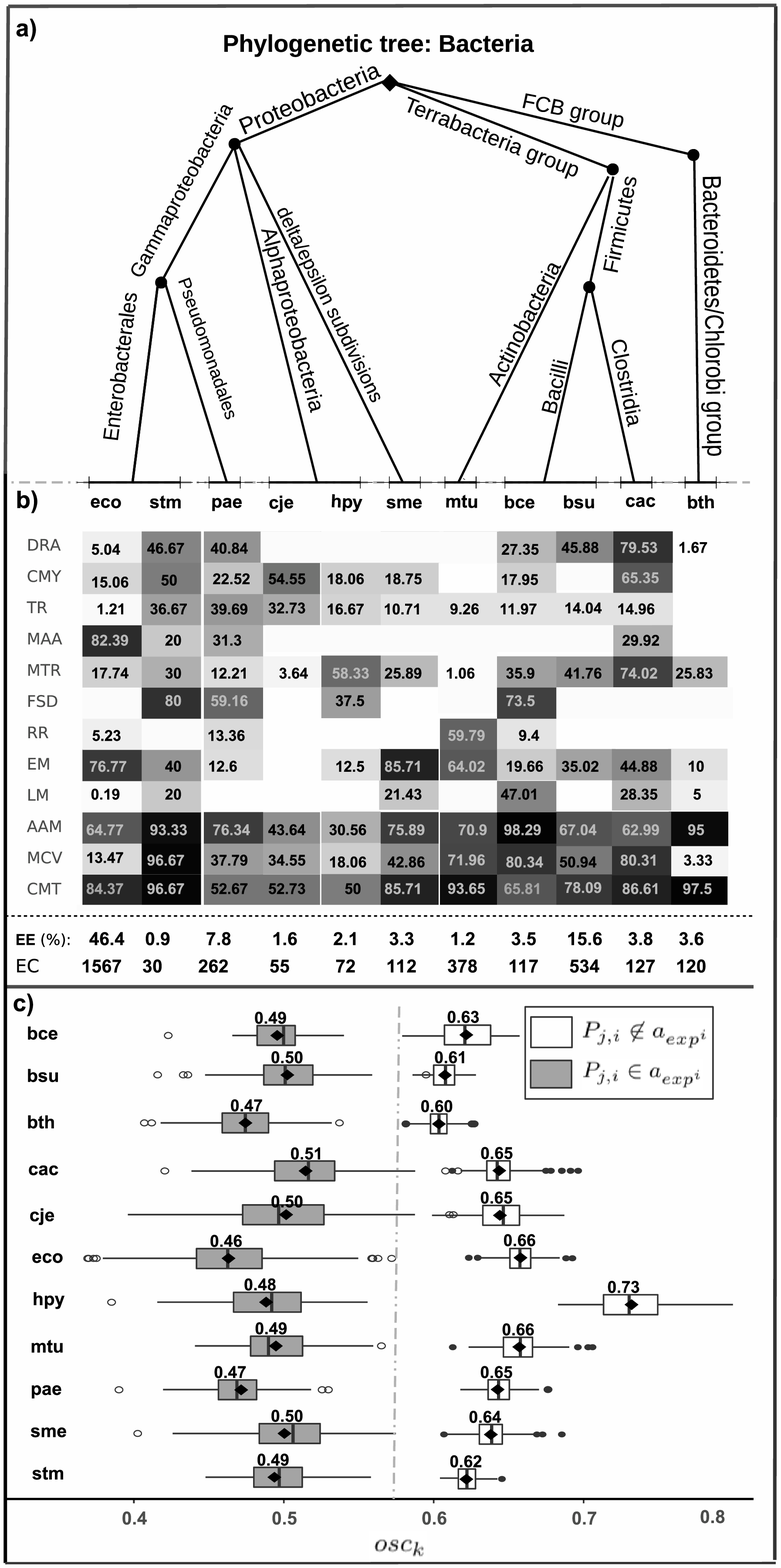}}
			\caption{ In this Figure, in box \textbf{a}, a phylogenetic tree is projected onto the heatmap between-organims and \textbf{KO Level 2} functionalities. In box \textbf{b} the row labels are specified as in Figure \ref{Fig3}, while the column labels are the organisms listed in Section \ref{subs2}.  Under the heatmap the cardinality of the experiments (\textbf{EC}) and their relative effort (\textbf{EE}) are shown in order to represent the relative influence of the oscillatory networks in phase synchronisations of Figure \ref{Fig3}. For example, E.\emph{coli} (eco) influences the $46\%$ of the intersections of Figure \ref{Fig3}. In Figure \ref{Fig5}, in box \textbf{c}, between-organisms boxplot comparisons with respect to the $osc_k$ distributions are shown. The phase synchronised oscillatory networks are represented by the gray boxplots and the inactive ones by the white boxplots. The plotted values represent their average $osc_k$ value.}\label{Fig5}
		\end{figure}
	
		\subsection{Multi-omic sources}
		\label{subs2}
		The dataset analysed in this work is composed of 2.830.722 multi-omic signals of 11 different bacteria on thousands of environmental experiments
		The bacteria included in the study are: \emph{Bacillus cereus} (ATCC 14579) [KEGG ID: \emph{bce}], \emph{Bacillus subtilis} (168) [KEGG ID: \emph{bsu}], \emph{Bacteroides thetaiotaomicron} (VPI-5482) [KEGG ID: \emph{bth}], \emph{Clostridium acetobutylicum}  (ATCC 824) [KEGG ID: \emph{cac}], \emph{Campylobacter jejuni} (NCTC 11168) [KEGG ID: \emph{cje}],  \emph{Escherichia coli} (K-12 MG1655) [KEGG ID: \emph{eco}],  \emph{Helicobacter pylori} (26695) [KEGG ID: \emph{hpy}],  \emph{Mycobacterium tuberculosis} (H37Rv) [KEGG ID: \emph{mtu}],  \emph{Pseudomonas aeruginosa} (PAO1) [KEGG ID: \emph{pae}],  \emph{Sinorhizobium meliloti} 1021 [KEGG ID: \emph{sme}] and  \emph{Salmonella enterica} (serovar Typhimurium LT2) [KEGG ID: \emph{stm}]. We define the set of these organisms as \\ $\mathbf{O} :=\lbrace bce, bsu, bth, cac,cje,eco,hpy,mtu, pae, sme, stm \rbrace$.
		The multi-omic signals were integrated for every organism ($\forall \mathbf{O}$) and for every $P_j$ following the combinations of single omic values $\vec{sv_{j,i}}$ with and without operon compression. The interplay between these omic layers is described by \cite{angione2016multiplex}. In particular, these multi-omic values ($\vec{mv_{j,i}}$) are combinations of dynamic $\vec{sv_{j,i}}$ and static $\vec{sv_{j,i}}$ or only a combination of static $\vec{sv_{j,i}}$. The static $\vec{sv_{j,i}}$ are the CAI (genomic layer) and the MW (proteomic layer). The $\vec{sv_{j,i}}$ of the CAI was computed as described by \cite{sharp1987codon}. The MW was computed considering the molecular weight of the amino-acidic composition of each protein in $P_j$. These values are called static because they do not change when perturbations occur. The dynamic $\vec{sv_{j,i}}$ were represented as ‘condition contrast’ (\emph{CC}) and represented the mRNA expression changes between microarray experiments. From a certain point of view, the \emph{CC} is a glue between the two static layers. The data for the condition contrast were downloaded from the COLOMBOS v3.0 dataset and they were already normalised within and between experiments by the creators of the dataset \citep{meysman2013colombos}. For this reason, through the \emph{CC} was possible to compare the between organisms signals $s \forall \mathbf{O}$ based on multi-platform experimental setups without losing generalisation. The four multi-omic combinations considered in this work were: $\mathbf{MOC}:=\lbrace(CC,CAI),(CC,MW),(CAI,MW),(CAI,CC,MW)\rbrace$.\\ The molecular networks pathway information and their KEGG Orthology are extracted from KEGG through a REST service \citep{kanehisa2000kegg, mao2005automated, kanehisa2013data}. The information about the operons localisation was determined through the OperonDB dataset \citep{pertea2008operondb}. The order of the genes was obtained through the NCBI dataset \citep{barrett2008ncbi} and aligned to the KEGG microbial genome information.  The whole dataset, with their respective labels, is provided in an R data format in the Supplementary Materials Section 6. The phylogenetic tree is reconstructed considering the NCBI taxonomy dataset \cite{federhen2011ncbi}.
		

	\section{Results and Discussion}
	\label{resdis} \textbf{I) Multi-omic signal oscillations}: The  mean absolute error (MAE) is computed in order to quantify the distance of $osc_s$ obtained from the original signals and the noised/perturbed ones (see also Supplementary Material - Section 4) \citep{ruckdeschel8distrmod} .
	The two types of perturbations adopted here are the same described in paragraph \ref{robsen}. MAE is computed for all the $s \in \mathbf{O}$ ( see Section \ref{subs2}) with $osc_s \geq \phi$ with $ \phi = 0.7$. Algorithm \ref{alg:algo2} computes the $osc_s$ of the original signal and of the noisy one in the same way.
	A cutoff  is applied to the wavelength, considering that the biological meaning of sequences makes sense for signals with $n \geq  6$. Nevertheless, very long signals present low oscillation indices with respect to median wavelengths, thus, respecting the \emph{short memory property} of \cite{ron1996power}. With these constraints, we are still considering the $ 95\% $ of the pathways and $63\%$ of pathways with operon compression. The oscillation index $osc_s$ is computed on signals with and without operon compression considering all the possible \textbf{MOC} separately. The MAE between the original signals and the perturbed ones is shown in Table \ref{Tab:02}. As it is shown, the distance between the noisy signals and the original ones is more pronounced considering the combination of CAI, molecular weight and condition contrasts (bold cells). Instead, the distance between the random shuffled signals and the original ones is more pronounced in the combination of CAI and MW (bold cells). In this case, the involved omics are both static and deeply related to the gene order.	In these analyses, it is highlighted that the multi-omic signals preserve the oscillation behaviours, described by our oscillation indices, representing effectively recurrent patterns present in nature. As it is shown also in Figure \ref{Fig4}), slight or massive multi-omic variations lower dramatically the oscillation index $osc_s$.  \\
	\textbf{II) Network-level synchronisations}: The synchronised signals, for each experiment, could give a meaningful picture of the network-level interplay of multi-omics. With the methodology described in section \ref{netsync}, important clues to support the hypothesis that there are groups of pathways/oscillatory networks in synchronisation are provided.
	For example, in Figure \ref{Fig3}, the most frequent network-level interplay based on  multi-omic signals \textbf{CAI-MW-CC}  for the functional feature of \textbf{KO Level 2} is shown. The intersections reported in Figure \ref{Fig3} involve the pathway signals of all 11 bacteria, shaping the general behavior of the dynamics of oscillations at the network level and outlining a similar response to several experimental perturbations.  According to \cite{fuhrer2005experimental}, if we look at, for example, Figure \ref{Fig5} box \textbf{(a-b)} we can see the central role of \emph{Carbohydrate Metabolism (CMT)}. In particular, the oscillatory networks belonging to the CMT class are in synchronisation in practically all the experiments and all the organisms (Figure \ref{Fig3} - see CMT row). Under the dictates of evolutionary pressure, in Figure \ref{Fig5} box \textbf{(a-b)}, it is possible to see the percentages of reciprocal influence that the signals have in synchronisation also for other important functions: \emph{Drug Resistance} (DRA), \emph{Cellular Motility} (CMY), etc.  Moreover, we investigated the behaviour of the oscillation index $osc_k$ computed for each organism for each pathway with \textbf{KO level 1} (Figure \ref{Fig5}, box \textbf{c}). The distributions of the $osc_k$ were studied by separating the synchronised signals (gray boxplots) to those with a lower oscillation behaviour (white boxplots). In this case, the oscillation index $osc_k$ of the synchronised signals is lower than in those not synchronised even if they show a higher oscillation index $osc_s$. It is evident that the two distributions are separable and preserved along with the organisms, except in rare cases where some few oscillatory networks seem to overlap the not synchronised ones. In these particular cases, we observed a higher $osc_k$ for these functional classes:  \emph{Drug Resistance, Cell Motility, Xenobiotic Metabolism, DNA Repair, Amino Acid Metabolism}. Instead, although the high values of $osc_s$ in Table \ref{Tab:02}, the distances between the $osc_k$ are considerably reduced on network-level oscillations of signals with operon compressions. This evidence could allow us to assume that network level synchronisations have a significant oscillation in amplitude with longer periods due to the underlying synchronisation mechanism that can be derived from the regulatory and control circuits. Conversely, a high $osc_k$ could indicate that networks show a rapid biological response to external stress. However, these hypotheses will have to be explored in future research work.
	\label{results}
		\begin{table}[!t]
		\begin{tabular}{lllll}
			\multicolumn{4}{l}{ }         &                         \\
			\multicolumn{4}{l}{ MAE on all the possible multi-omic combinations (\textbf{MOC}):} & \\
			\textbf{$osc_s$} & \emph{$osc_s$ vs noise} & \emph{$osc_s$ vs shuffle} & \emph{Signals set-size} & \emph{S-s \%} \\
			\textbf{CAI-MW-CC}            & \textbf{0.81}             & 0.72                    & 188649 on 655272     & 29\%\\
			MW-CC                         & 0.74                      & 0.73                    & 211725 on  65527     & 32\%\\
			CAI-CC                        & 0.77                      & 0.67                    & 177349 on  65527     & 27\%\\
			\textbf{CAI-MW}               & 0.80                      & \textbf{0.81}           & 204 on 779           & 26\%\\
			\multicolumn{4}{l}{ }         &                         \\
			\multicolumn{4}{l}{MAE with all the possible \textbf{MOC} with operon compression:} & \\
			\textbf{$osc_s$} & \emph{$osc_s$ vs noise} & \emph{$osc_s$ vs shuffle} & \emph{Signals set-size} & \emph{S-s \%} \\
			\textbf{CAI-MW-CC}            & \textbf{0.93}             & 0.82                    & 58350 on 287933     & 20\%\\
			MW-CC                         & 0.86                      & 0.84                    & 67384 on 287933     & 23\%\\
			CAI-CC                        & 0.90                      & 0.73                    & 72911 on 287933     & 25\%\\
			\textbf{CAI-MW}               & 0.91                      & \textbf{0.85}           & 73 on  328          & 22\% \\
		\end{tabular}
		\caption{\label{Tab:02}:   Table of distances between the original $osc_s$ and the perturbed ones. In Table \ref{Tab:02} the mean absolute error (MAE) between the original signal oscillation indices ($osc_s \geq \phi$, with $\phi = 0.7$) and the perturbed ones are shown. The additive noise and the random shuffle are the two perturbations listed as \emph{noise} and \emph{shuffle}. The distances are computed for each \textbf{MOC}. In \emph{Signals set-size}, the size of the signals with the in-text described constraints is reported.  Despite the set-size that remains equal for each organism, every pathway and every MOC, the number of multi-omic operon compressed signals is different from the original ones because some pathways during the operon compression became too short to be considered as biological sequences.}

	\end{table}
	\section{Conclusion}
	In this article, two new signal metrics have been introduced to study multi-omic oscillations at the network level and on single-snapshot experiments, defined in particular as condition contrasts. From the analysis of these metrics, it is possible to provide interesting clues about the characteristics of the signal proving that multi-omic network-level oscillations exist in nature. Furthermore, clear evidence has been provided that these oscillations could show a synchronised interaction in response to perturbations. Multi-omic signal analyses have been extended to multiple organisms and related to their phylogenetic tree to provide better comparisons. Algorithmic methodologies are provided and accompanied by a tool on supplementary material and on the online repository. This work could be useful in the fields of synthetic biology and systems biology with the goal of mapping the organism regulation and control circuits, for example, in case of lack of time series experiments. Furthermore, there is a growing amount of whole-genome and longitudinal data and these metrics answer to the need to detect complex patterns of changes.

	\bibliographystyle{plain}
	\bibliography{main.bib}

\begin{thebibliography}{}

\bibitem[Amariei {\em et~al.}(2014)Amariei, Tomita, and
  Murray]{amariei2014quantifying}
Amariei, C.  {\em et~al.} (2014).
\newblock Quantifying periodicity in omics data.
\newblock {\em Frontiers in cell and developmental biology\/}, {\bf 2}, 40.

\bibitem[Angione {\em et~al.}(2016)Angione, Conway, and
  Li{\'o}]{angione2016multiplex}
Angione, C.  {\em et~al.} (2016).
\newblock Multiplex methods provide effective integration of multi-omic data in
  genome-scale models.
\newblock {\em BMC bioinformatics\/}, {\bf 17}(4), 83.

\bibitem[Arenas {\em et~al.}(2008)Arenas, D{\'\i}az-Guilera, Kurths, Moreno,
  and Zhou]{arenas2008synchronization}
Arenas, A.  {\em et~al.} (2008).
\newblock Synchronization in complex networks.
\newblock {\em Physics reports\/}, {\bf 469}(3), 93--153.

\bibitem[Bardozzo {\em et~al.}(2015)Bardozzo, Li{\'o}, and
  Tagliaferri]{bardozzo2015multi}
Bardozzo, F.  {\em et~al.} (2015).
\newblock Multi omic oscillations in bacterial pathways.
\newblock In {\em 2015 International Joint Conference on Neural Networks
  (IJCNN)\/}, pages 1--8. IEEE.

\bibitem[Bardozzo {\em et~al.}(2018)Bardozzo, Li{\'o}, and
  Tagliaferri]{bardozzo2018study}
Bardozzo, F.  {\em et~al.} (2018).
\newblock A study on multi-omic oscillations in escherichia coli metabolic
  networks.
\newblock {\em BMC bioinformatics\/}, {\bf 19}(7), 194.

\bibitem[Barrett {\em et~al.}(2008)Barrett, Troup, Wilhite, Ledoux, Rudnev,
  Evangelista, Kim, Soboleva, Tomashevsky, Marshall, {\em
  et~al.}]{barrett2008ncbi}
Barrett, T.  {\em et~al.} (2008).
\newblock Ncbi geo: archive for high-throughput functional genomic data.
\newblock {\em Nucleic acids research\/}, {\bf 37}(suppl\_1), D885--D890.

\bibitem[Boukouvalas {\em et~al.}(2019)Boukouvalas, Cutillo, Marinopoulou,
  Papalopulu, and Rattray]{boukouvalas2019osconet}
Boukouvalas, A.  {\em et~al.} (2019).
\newblock Osconet: Inferring oscillatory gene networks.
\newblock {\em bioRxiv\/}, page 600049.

\bibitem[Darkhovski(1994)Darkhovski]{darkhovski1994nonparametric}
Darkhovski, B.~S. (1994).
\newblock Nonparametric methods in change-point problems: A general approach
  and some concrete algorithms.
\newblock {\em Lecture Notes-Monograph Series\/}, pages 99--107.

\bibitem[Elowitz and Leibler(2000)Elowitz and Leibler]{elowitz2000synthetic}
Elowitz, M.~B. and Leibler, S. (2000).
\newblock A synthetic oscillatory network of transcriptional regulators.
\newblock {\em Nature\/}, {\bf 403}(6767), 335--338.

\bibitem[Federhen(2011)Federhen]{federhen2011ncbi}
Federhen, S. (2011).
\newblock The ncbi taxonomy database.
\newblock {\em Nucleic acids research\/}, {\bf 40}(D1), D136--D143.

\bibitem[Fuhrer {\em et~al.}(2005)Fuhrer, Fischer, and
  Sauer]{fuhrer2005experimental}
Fuhrer, T.  {\em et~al.} (2005).
\newblock Experimental identification and quantification of glucose metabolism
  in seven bacterial species.
\newblock {\em Journal of bacteriology\/}, {\bf 187}(5), 1581--1590.

\bibitem[Golden(2003)Golden]{golden2003timekeeping}
Golden, S.~S. (2003).
\newblock Timekeeping in bacteria: the cyanobacterial circadian clock.
\newblock {\em Current opinion in microbiology\/}, {\bf 6}(6), 535--540.

\bibitem[Govindarajan {\em et~al.}(2012)Govindarajan, Nevo-Dinur, and
  Amster-Choder]{govindarajan2012compartmentalization}
Govindarajan, S.  {\em et~al.} (2012).
\newblock Compartmentalization and spatiotemporal organization of
  macromolecules in bacteria.
\newblock {\em FEMS microbiology reviews\/}, {\bf 36}(5), 1005--1022.

\bibitem[Guantes {\em et~al.}(2010)Guantes, Estrada, and
  Poyatos]{guantes2010trade}
Guantes, R.  {\em et~al.} (2010).
\newblock Trade-offs and noise tolerance in signal detection by genetic
  circuits.
\newblock {\em PLoS One\/}, {\bf 5}(8), e12314.

\bibitem[Hawe {\em et~al.}(2019)Hawe, Theis, and Heinig]{hawe2019inferring}
Hawe, J.~S.  {\em et~al.} (2019).
\newblock Inferring interaction networks from multi-omics data.
\newblock {\em Frontiers in genetics\/}, {\bf 10}, 535.

\bibitem[Kanehisa and Goto(2000)Kanehisa and Goto]{kanehisa2000kegg}
Kanehisa, M. and Goto, S. (2000).
\newblock Kegg: kyoto encyclopedia of genes and genomes.
\newblock {\em Nucleic acids research\/}, {\bf 28}(1), 27--30.

\bibitem[Kanehisa {\em et~al.}(2013)Kanehisa, Goto, Sato, Kawashima, Furumichi,
  and Tanabe]{kanehisa2013data}
Kanehisa, M.  {\em et~al.} (2013).
\newblock Data, information, knowledge and principle: back to metabolism in
  kegg.
\newblock {\em Nucleic acids research\/}, {\bf 42}(D1), D199--D205.

\bibitem[Lenz and S{\o}gaard-Andersen(2011)Lenz and
  S{\o}gaard-Andersen]{lenz2011temporal}
Lenz, P. and S{\o}gaard-Andersen, L. (2011).
\newblock Temporal and spatial oscillations in bacteria.
\newblock {\em Nature Reviews Microbiology\/}, {\bf 9}(8), 565.

\bibitem[Levine {\em et~al.}(2013)Levine, Lin, and
  Elowitz]{levine2013functional}
Levine, J.~H.  {\em et~al.} (2013).
\newblock Functional roles of pulsing in genetic circuits.
\newblock {\em Science\/}, {\bf 342}(6163), 1193--1200.

\bibitem[Lutkenhaus(2008)Lutkenhaus]{lutkenhaus2008min}
Lutkenhaus, J. (2008).
\newblock Min oscillation in bacteria.
\newblock In {\em Cellular Oscillatory Mechanisms\/}, pages 49--61. Springer.

\bibitem[Mao {\em et~al.}(2005)Mao, Cai, Olyarchuk, and Wei]{mao2005automated}
Mao, X.  {\em et~al.} (2005).
\newblock Automated genome annotation and pathway identification using the kegg
  orthology (ko) as a controlled vocabulary.
\newblock {\em Bioinformatics\/}, {\bf 21}(19), 3787--3793.

\bibitem[Meysman {\em et~al.}(2013)Meysman, Sonego, Bianco, Fu,
  Ledezma-Tejeida, Gama-Castro, Liebens, Michiels, Laukens, Marchal, {\em
  et~al.}]{meysman2013colombos}
Meysman, P.  {\em et~al.} (2013).
\newblock Colombos v2. 0: an ever expanding collection of bacterial expression
  compendia.
\newblock {\em Nucleic acids research\/}, {\bf 42}(D1), D649--D653.

\bibitem[Michalodimitrakis and Isalan(2008)Michalodimitrakis and
  Isalan]{michalodimitrakis2008engineering}
Michalodimitrakis, K. and Isalan, M. (2008).
\newblock Engineering prokaryotic gene circuits.
\newblock {\em FEMS microbiology reviews\/}, {\bf 33}(1), 27--37.

\bibitem[Pertea {\em et~al.}(2008)Pertea, Ayanbule, Smedinghoff, and
  Salzberg]{pertea2008operondb}
Pertea, M.  {\em et~al.} (2008).
\newblock Operondb: a comprehensive database of predicted operons in microbial
  genomes.
\newblock {\em Nucleic acids research\/}, {\bf 37}(suppl\_1), D479--D482.

\bibitem[Potvin-Trottier {\em et~al.}(2016)Potvin-Trottier, Lord, Vinnicombe,
  and Paulsson]{potvin2016synchronous}
Potvin-Trottier, L.  {\em et~al.} (2016).
\newblock Synchronous long-term oscillations in a synthetic gene circuit.
\newblock {\em Nature\/}, {\bf 538}(7626), 514--517.

\bibitem[Prokop and Csuk{\'a}s(2013)Prokop and Csuk{\'a}s]{prokop2013systems}
Prokop, A. and Csuk{\'a}s, B. (2013).
\newblock {\em Systems biology: integrative biology and simulation tools\/},
  volume~1.
\newblock Springer Science \& Business Media.

\bibitem[{R}(2018){R}]{Rrrr}
{R} (2018).
\newblock {\em R: A Language and Environment for Statistical Computing\/}.
\newblock R Foundation for Statistical Computing, Vienna, Austria.

\bibitem[Razali {\em et~al.}(2011)Razali, Wah, {\em et~al.}]{razali2011power}
Razali, N.~M.  {\em et~al.} (2011).
\newblock Power comparisons of shapiro-wilk, kolmogorov-smirnov, lilliefors and
  anderson-darling tests.
\newblock {\em Journal of statistical modeling and analytics\/}, {\bf 2}(1),
  21--33.

\bibitem[Rocha(2003)Rocha]{rocha2003dna}
Rocha, E.~P. (2003).
\newblock Dna repeats lead to the accelerated loss of gene order in bacteria.
\newblock {\em TRENDS in Genetics\/}, {\bf 19}(11), 600--603.

\bibitem[Ron {\em et~al.}(1996)Ron, Singer, and Tishby]{ron1996power}
Ron, D.  {\em et~al.} (1996).
\newblock The power of amnesia: Learning probabilistic automata with variable
  memory length.
\newblock {\em Machine learning\/}, {\bf 25}(2-3), 117--149.

\bibitem[Ruckdeschel and Kohl(2018)Ruckdeschel and Kohl]{ruckdeschel8distrmod}
Ruckdeschel, P. and Kohl, M. (2018).
\newblock distrmod—an s4-class based package for statistical models.
\newblock {\em Robust Inference in Generalized Linear Models\/}, {\bf 10}, 159.

\bibitem[Sharp and Li(1987)Sharp and Li]{sharp1987codon}
Sharp, P.~M. and Li, W.-H. (1987).
\newblock The codon adaptation index-a measure of directional synonymous codon
  usage bias, and its potential applications.
\newblock {\em Nucleic acids research\/}, {\bf 15}(3), 1281--1295.

\bibitem[Shis {\em et~al.}(2018)Shis, Bennett, and Igoshin]{shis2018dynamics}
Shis, D.~L.  {\em et~al.} (2018).
\newblock Dynamics of bacterial gene regulatory networks.
\newblock {\em Annual review of biophysics\/}, {\bf 47}, 447--467.

\bibitem[Siegmund(2013)Siegmund]{siegmund2013change}
Siegmund, D. (2013).
\newblock Change-points: from sequential detection to biology and back.
\newblock {\em Sequential analysis\/}, {\bf 32}(1), 2--14.

\bibitem[Singh and Soltani(2013)Singh and Soltani]{singh2013quantifying}
Singh, A. and Soltani, M. (2013).
\newblock Quantifying intrinsic and extrinsic variability in stochastic gene
  expression models.
\newblock {\em Plos one\/}, {\bf 8}(12), e84301.

\bibitem[Tamames(2001)Tamames]{tmes2001evolution}
Tamames, J. (2001).
\newblock Evolution of gene order conservation in prokaryotes.
\newblock {\em Genome biology\/}, {\bf 2}(6), research0020--1.

\bibitem[Unakafov and Keller(2018)Unakafov and Keller]{unakafov2018change}
Unakafov, A. and Keller, K. (2018).
\newblock Change-point detection using the conditional entropy of ordinal
  patterns.
\newblock {\em Entropy\/}, {\bf 20}(9), 709.

\bibitem[Venables and Ripley(2013)Venables and Ripley]{venables2013modern}
Venables, W.~N. and Ripley, B.~D. (2013).
\newblock {\em Modern applied statistics with S-PLUS\/}.
\newblock Springer Science \& Business Media.

\bibitem[Wang and Levin(2009)Wang and Levin]{wang2009metabolism}
Wang, J.~D. and Levin, P.~A. (2009).
\newblock Metabolism, cell growth and the bacterial cell cycle.
\newblock {\em Nature Reviews Microbiology\/}, {\bf 7}(11), 822--827.

\end{thebibliography}
\end{document}